\documentclass[12pt]{article}
\usepackage{amsmath,amssymb}
\usepackage{graphicx,color}
\numberwithin{equation}{section}
\usepackage{cite}
\usepackage{bm}
\usepackage{dcolumn}
\def\bea{\begin{eqnarray}}
\def\eea{\end{eqnarray}}

\newcommand{\nn}{\nonumber}
\newcommand{\na}{\nabla}
\def\beq{\begin{equation}}
\def\eeq{\end{equation}}

\def\na{\nabla}

\def\M{\mu}
\def\N{\nu}

\begin{document}

\vspace{12mm}

\begin{center}
{{{\Large {\bf Spontaneous symmetry breaking in 5D conformally invariant gravity}}}}\\[10mm]

{Taeyoon Moon$^{a,b}$\footnote{e-mail address: dpproject@skku.edu}
and Phillial Oh$^{a}$\footnote{e-mail address: ploh@skku.edu}
}\\[8mm]

{\em {${}^{a}$ Department of Physics and Institute of Basic Science,\\
Sungkyunkwan University, Suwon 440-746 Korea\\[7pt]
${}^{b}$ Institute of Basic Science and Department of Computer
Simulation, Inje University,\\
Gimhae 621-749, Korea}\\[7pt]
}
\end{center}
\vspace{2mm}

\begin{abstract}
We explore the possibility of the spontaneous symmetry breaking in
5D conformally invariant gravity, whose action consists of a scalar
field nonminimally coupled to the curvature with its potential.
Performing dimensional reduction via ADM decomposition, we find that
the model allows an exact solution giving rise to the 4D Minkowski
vacuum. Exploiting the conformal invariance with Gaussian warp
factor, we show that it also admits a solution which implement the
spontaneous breaking of conformal symmetry. We investigate its
stability by performing the tensor perturbation and find the
resulting system is described by the conformal quantum mechanics.
Possible applications to the spontaneous symmetry breaking of
time-translational symmetry along the dynamical fifth direction and
the brane-world scenario are discussed.
\end{abstract}
\vspace{5mm}


\vspace{1.5cm}

\hspace{11.5cm}{Typeset Using \LaTeX}
\newpage
\renewcommand{\thefootnote}{\arabic{footnote}}
\setcounter{footnote}{0}
\section{Introduction}

Conformal symmetry is an important idea which has appeared in
diverse area of physics, and its application to gravity has started
with the idea that
 conformally invariant gravity in four dimensions (4D) \cite{Weyl:1918ib}
 might result in a unified description of gravity and electromagnetism.
 The Einstein-Hilbert action of general relativity is not conformally invariant.
 In realizing the conformal invariance of this, a conformal scalar field is
 necessary \cite{Dirac:1938mt} in order to compensate the conformal
 transformation of the metric, and a quartic potential for the scalar field
 can be allowed. Its higher dimensional extensions are straightforward.
 In five dimensions (5D), conformal symmetry can be preserved with a fractional
 power potential \cite{Jackiw:2011vz} for the scalar field. So far, it seems
 that little attention has been paid to the 5D conformal gravity with
 its fractional power potential. Such a potential renders  a perturbative
 approach inaccessible,
 but non-perturbative treatment may reveal novel aspects. One can also
 construct a conformally invariant gravity with the Weyl tensor via $R$-squared
 gravity, but we focus on the Einstein gravity with a conformal scalar.

 If the scalar field spontaneously breaks the conformal invariance with a
 Planckian VEV, the theory reduces to the 4D Einstein gravity with
 a cosmological constant \cite{Padmanabhan:1986yc,Demir:2004kc}.
 On the other hand, the spontaneous symmetry breaking of Lorentz
 symmetry \cite{Bertolami:2006bf} or gauge symmetry \cite{Bertolami:2007dt}
 in 5D brane-world scenario \cite{ArkaniHamed:1998rs}
was studied, but little is known in the context
 of 5D conformal gravity. In this paper, we explore 5D conformal gravity
 with a conformal scalar and investigate possible consequences in view of
 the spontaneous symmetry breaking.

Let us consider 5D conformal scalar action of the form \bea S=\int
d^{5}x \sqrt{-g} \left[
 \frac{1}{2} \xi \phi^2R
 -\frac{\omega}{2}g^{AB}D_{A}{\phi}D_{B}\phi-V(\phi) \right]
+S_{m}, \label{conac}  \eea where $R$ is the five dimensional
curvature scalar, $S_m$ is the action for some matter\footnote{We
assume that the matter is confined on a hypersurface at $y=y_m$,
where $y$ is the fifth coordinate. It is known that this matter in
the brane has a geometrical origin in space-time-matter (STM) theory
or induced-matter theory (IMT)
\cite{Wesson:1999nq,PonceDeLeon:2001un,Doroud:2009zza}. Also, the
IMT has been extended to the modified Brans-Dicke theory (MBDT) of
the type (\ref{conac}), where the induced matter exhibits
interesting cosmological consequences
\cite{Rasouli:2014dxa,Rasouli:2016ngl}. However, in this paper we
are interested in the phenomena of spontaneous symmetry breaking and
thus, we will be neglecting the matter sector.}, and $A,B$ run over
$0,1,2,3,4$. Here, $\xi$ is a dimensionless parameter describing the
nonminimal coupling of the scalar field to the spacetime curvature.
Also a parameter $\omega=\pm 1$ with $+(-)$ corresponds to
canonical(ghost) scalar. For $\omega=\pm1$, the conformal invariance
of the action (\ref{conac}) without matter term forces $\xi$ to be
\begin{eqnarray}\label{xiv}
V(\phi)=V_0|\phi|^{\frac{10}{3}}, ~ \xi=\mp\frac{3}{16},\label{s3}
\end{eqnarray}
where $V_0$ is a constant and the corresponding conformal transformation is given by
\begin{eqnarray}\label{cont}
g_{AB}\rightarrow e^{2\sigma(x)}g_{AB},~\phi\rightarrow
e^{-\frac{3}{2}\sigma(x)}\phi(x).
\end{eqnarray}
When $\omega=-1$, the conformal scalar has a negative kinetic energy
term, but we regard it as a gauge artifact \cite{Bars:2013yba} which
can be eliminated from the beginning through field redefinition.
Even with no scalar field remaining after gauging away for both
cases ($\omega=\pm1$), the physical mass scale can be set since the
corresponding vacuum solution requires introduction of a scale which
characterizes the conformal symmetry breaking. In 4D conformal
gravity, it is known that the conformal symmetry can be
spontaneously broken at electroweak \cite{Cheng:1988zx,Demir:2004kc}
or Planck \cite{Padmanabhan:1986yc,Demir:2004kc} scale. In all
cases, the action (\ref{conac}) becomes 5D Einstein action with a
cosmological constant by redefinition of the metric, $\hat
g_{AB}=\phi^{\frac{4}{3}}g_{AB}$, but we stick to the above
conformal form (\ref{conac}) of the action to argue with the
spontaneous breakdown of the conformal symmetry.

The paper is organized as follows. In Sec. 2, we perform the
dimensional reduction from five to four dimension by using the ADM
decomposition. In Sec. 3, we present exact solutions with four
dimensional Minkowski vacuum ($R_{(4)}=0$) and check if they can
give a spontaneous breaking of the conformal symmetry. In Sec. 4,
the gravitational perturbation and  their stability for the solutions are considered.  In Sec. 5, we includes  the summary  and discussions.

\section{Dimensional reduction ($5D$ to $4D$)}

In order to derive the 4D action from the 5D conformal gravity
(\ref{conac}), we  make use of the following ADM decomposition where
the metric in $5D$ can be written as\footnote{$x^\mu$ are the
coordinates in $4D$ and $y$ is the non-compact coordinate along the
extra dimension (see \cite{Rasouli:2014dxa,Rasouli:2016ngl} and
references therein). We use spacetime signature $(-, +, +,+, \pm)$,
while $\epsilon = \pm 1$ denotes spacelike or timelike extra
dimension. }
\begin{equation}
dS^2 = g_{AB}d x^A d x^B = g_{\mu\nu}(x, y)(d x^{\mu} +N^{\mu} dy)(d
x^{\nu}+N^{\nu} dy) + \epsilon N^2(x, y) d y^2.\label{metric}
\end{equation}
To describe the background solution, we go to the ``comoving'' gauge
and choose $N^\mu=0.$
In this case, we can recover our $4D$ spacetime by going onto a hypersurface
$\Sigma_{y}:y =  y_{0} = $ constant, which is orthogonal to the $5D$ unit vector
\begin{equation}
\label{unit vector n}
{\hat{n}}^{A} = \frac{\delta^{A}_{4}}{N}, \;\;\;n_{A}n^{A} = \epsilon,
\end{equation}
along the extra dimension, and $g_{\mu\nu}$ can be interpreted as the metric of the 4D spacetime.
Using the metric ansatz (\ref{metric}), one obtains
\begin{eqnarray}
R^{(5)}=R^{(4)}-2\frac{ \nabla^2 N}{N}+\frac{\epsilon}{N^2}
 \left(\frac{\stackrel{\ast}N g^{\alpha\beta}\stackrel{\ast}g_{\alpha\beta}}{N} - g^{\alpha\beta} \stackrel{\ast \ast}g_{\alpha\beta}-
\frac{  3\stackrel{\ast}  g^{\alpha\beta} \stackrel{\ast} g_{\alpha\beta}}{4} - \frac{( g^{\alpha\beta}\stackrel{\ast}g_{\alpha\beta})^2}{4}\right),
\end{eqnarray}
where the asterisk ${}^*$ denotes the differentiation with respect to $y$ and $\nabla^2=\nabla_{\mu}\nabla^{\mu}$
is the four dimensional Laplacian.
Using this, we find the action (\ref{conac}) becomes
\begin{eqnarray}\label{react}
S&=&\int d^{5}x \sqrt{-g^{(4)}}N\Bigg[
 \frac{1}{2} \xi \phi^2R^{(4)}+\epsilon\frac{\xi \phi^2}{8N^2}
 \left(   \stackrel{\ast}  g^{\alpha\beta}
 \stackrel{\ast} g_{\alpha\beta}+( g^{\alpha\beta}\stackrel{\ast}g_{\alpha\beta})^2
+\frac{8\stackrel{\ast}\phi}{\phi} g^{\alpha\beta} \stackrel{\ast}g_{\alpha\beta} \right)\nn\\
&&\hspace*{6em}
-\xi\nabla^2\phi^2\ -\frac{\omega}{2}g^{\M\N}\na_{\M}{\phi}\na_{\N}\phi-
 \epsilon\frac{\omega}{2N^2}
 \stackrel{\ast}\phi\stackrel{\ast}\phi-V(\phi)\Bigg]
+S_{m}.
\end{eqnarray}
One can check that the above action (\ref{react}) is invariant with respect to four dimensional diffeomorphism
$x^\mu\rightarrow x^{\prime \mu}\equiv x^{\prime \mu}(x)$
with $N^\prime (x^\prime, y)=N(x, y).$
It is also invariant under $y\rightarrow y^\prime\equiv y^\prime(y)$ and $N\rightarrow N^\prime\equiv (dy^\prime/dy)^{-1}N$
apart from the matter action.

Before going further, we would like to comment on the homogeneous solution to the equations of motion given in
5D conformal gravity (\ref{conac}) without matter term. To this end, we  first consider the Einstein equation for the action (\ref{conac}),
whose form is given by
\bea
&&\hspace*{5em}R_{AB}-\frac{1}{2}g_{AB}R~=~T_{AB},\label{Gmn}\\
T_{AB}&=& \frac{1}{\xi\phi^2}\left(\partial_A\phi \partial_B\phi-
\frac{1}{2}g_{AB}\partial^C\phi\partial_C\phi-Vg_{AB}\right
)+\frac{1}{\phi^2} \left( D_A D_B
\phi^2-g_{AB}\square^{(5)}\phi^2\right) \eea and the scalar equation
can be written as \bea 0=\omega\square^{(5)}\phi+\xi\phi
R-V^\prime(\phi),\label{moe} \eea where $D_A$ is 5D covariant
derivative, $\square^{(5)}\equiv D_A D^A$, and the prime ${}^{'}$
denotes the differentiation with respect to $\phi$. One can easily
check that the homogeneous solution to Eqs.(\ref{Gmn}) and
(\ref{moe}) is given by
\begin{eqnarray}\label{homog}
R_{AB}=\Lambda g_{AB},~~\phi=\phi_0, ~~\Lambda =\frac{2V_0}{3\xi}(\phi_0)^{4/3}.
\end{eqnarray}
We note that this solution can be approached in diverse ways.
Firstly,  field redefinition $\tilde
g_{MN}=(\phi/\phi_0)^{4/3}g_{MN}$ necessitate introduction of scale,
which leads to five dimensional Planck mass $M_5=\xi\phi_0^{2/3}$.
Secondly, the solution (\ref{homog}) corresponds to a gauge fixed
case ($\phi=\phi_0$) through the conformal transformation
(\ref{cont}). In the last, it can be interpreted as a vacuum
solution obtained when considering an effective potential $V_{\rm
eff}=-\xi R\phi^2/2+V_0|\phi|^{10/3}$ (we shall study the effective
potential $V_{\rm eff}$ for details at the end of the next section).
In all cases, conformal symmetry is spontaneously broken with a
symmetry breaking scale $\sim\phi_0\neq0$. In addition, they yield
the physically equivalent results: de Sitter $V_0/\xi>0$ or anti-de
Sitter $V_0/\xi<0$. It is to be noticed that both cases are
classically stable. Also there is a huge degeneracy of vacuum
solutions due to conformal invariance such that if
$\Big(g^{(0)}_{AB}(x,y), \phi^{(0)}(x,y)$\Big) is a solution, then
$\Big(\tilde g_{AB}=e^{2\sigma(x,y)}g^{(0)}_{AB},
\tilde{\phi}=e^{-\frac{3}{2}\sigma(x,y)} \phi^{(0)}\Big)$ is also a
solution for an arbitrary function $\sigma(x,y)$. In the next
section, we will investigate the explicit solution form, starting
from the reduced action (\ref{react}) without matter term.

\section{Exact solutions}

From the action (\ref{react}) without matter term,
 we find the equation of motion for $N$ as
\begin{eqnarray}
0&=&\frac{\xi \phi^2}{2} R^{(4)}-\frac{\epsilon\xi \phi^2}{8N^2}
 \left(   \stackrel{\ast}  g^{\alpha\beta}
 \stackrel{\ast} g_{\alpha\beta}+(g^{\mu\nu}\stackrel{\ast }
g_{\mu\nu})^2
+\frac{8\stackrel{\ast}\phi}{\phi} g^{\alpha\beta} \stackrel{\ast}g_{\alpha\beta} \right)
-\xi\nabla^2\phi^2\ \nonumber\\
&&\hspace*{11em}-\frac{\omega}{2}g^{\M\N}\na_{\M}{\phi}\na_{\N}\phi+\frac{
 \epsilon\omega}{2N^2}
 \stackrel{\ast}\phi\stackrel{\ast}\phi-V(\phi)\label{constraint1}
\end{eqnarray}
and the equation for four dimensional metric $g^{\mu\nu}$ is given by
\begin{eqnarray}
\frac{1}{2} \xi \phi^2 \Big(R_{\mu\nu}^{(4)}-\frac{1}{2}g_{\mu\nu}R^{(4)}\Big)
=T_{\mu\nu}^{(1)}+T_{\mu\nu}^{(2)}+T_{\mu\nu}^{(3)}+T_{\mu\nu}^{(4)}
+T_{\mu\nu}^{(5)},\label{einstein}
\end{eqnarray}
\begin{eqnarray}
T_{\mu\nu}^{(1)}&=&\frac{\xi}{2N}\left[\nabla_\mu\nabla_\nu(N\phi^2)
-g_{\mu\nu}\nabla^2(N\phi^2)\right],\nn\\
T_{\mu\nu}^{(2)}&=&\frac{\epsilon\xi\phi^2}{8N^2}\Bigg[
\frac{1}{2}\stackrel{\ast}  g^{\alpha\beta}
 \stackrel{\ast} g_{\alpha\beta}g_{\mu\nu}+2\frac{N}{\phi^2}
\stackrel{\ast}{\left(\frac{\phi^2}{N}\right)} \stackrel{\ast} g_{\mu\nu}
+2\stackrel{\ast\ast} g_{\mu\nu}+g^{\alpha\beta}\stackrel{\ast}g_{\alpha\beta}\stackrel{\ast} g_{\mu\nu}+\stackrel{\ast}  g^{\alpha\beta}g_{\beta\mu}\stackrel{\ast}g_{\nu\alpha}\nn\\
&&+ g^{\alpha\beta}\stackrel{\ast}g_{\beta\nu}\stackrel{\ast}g_{\mu\alpha} -\frac{1}{2}(g^{\alpha\beta}
\stackrel{\ast }g_{\alpha\beta})^2g_{\mu\nu}-
2\stackrel{\ast}{\left(g^{\alpha\beta}\stackrel{\ast}g_{\alpha\beta}\right)}g_{\mu\nu}
-2 \frac{N}{\phi^2}
\stackrel{\ast}{\left(\frac{\phi^2}{N}\right)}g^{\alpha\beta}\stackrel{\ast}g_{\alpha\beta}
g_{\mu\nu}\Bigg],\nn\\
T_{\mu\nu}^{(3)}&=&-\frac{\xi}{N}\left[\nabla_{(\mu}N\nabla_{\nu)}\phi^2
-\frac{1}{2}g_{\mu\nu}\nabla_\rho N\nabla^\rho\phi^2\right],\nn\\
T_{\mu\nu}^{(4)}&=&\left[\frac{\omega}{2}\nabla_{\mu}\phi\nabla_{\nu}\phi
-\frac{\omega}{4}g_{\mu\nu}\nabla_\rho\phi\nabla^\rho\phi-\frac{1}{2}V(\phi)g_{\mu\nu}\right],\nn\\
T_{\mu\nu}^{(5)}&=&-\frac{\epsilon}{N}\left[ {\xi}\stackrel{\ast}
{\left(\frac{\phi\stackrel{\ast}\phi}{N}\right)}g_{\mu\nu}+\frac{\omega}{4}
\frac{\stackrel{\ast}\phi\stackrel{\ast}\phi}{N} g_{\mu\nu} \right].
\end{eqnarray}
Also the equation of motion for the scalar field can be written as
\begin{eqnarray}
0&=&  \xi N \phi R^{(4)}-2\xi\phi\nabla^2N
+\frac{\epsilon\xi \phi}{4N}
 \left(   \stackrel{\ast}  g^{\alpha\beta}
 \stackrel{\ast} g_{\alpha\beta}-(g^{\alpha\beta}\stackrel{\ast }
g_{\alpha\beta})^2-4\stackrel{\ast}{\left(g^{\alpha\beta}\stackrel{\ast}g_{\alpha\beta}\right)}
+4\frac{\stackrel{\ast}N}{N}g^{\alpha\beta}\stackrel{\ast}g_{\alpha\beta}\right)
  \nn\\
&&+N\left(\omega\nabla_\mu\na^{\M}\phi+\frac{\omega}{N}\na_{\M}N\na^\M\phi-V^\prime
(\phi)\right)+\frac{\epsilon\omega}{2N}\left(g^{\alpha\beta}\stackrel{\ast}g_{\alpha\beta}  \stackrel{\ast}\phi
 -2\frac{\stackrel{\ast}N\stackrel{\ast}\phi}{N}+2
\stackrel{\ast\ast}\phi\right).\label{dpotential}
\end{eqnarray}
Now one can equate Eq. (\ref{constraint1}) and trace of
(\ref{einstein}), then obtain
\begin{eqnarray}
V(\phi)&=&-\frac{\xi}{2}\nabla^2\phi^2-2\frac{\xi}{N} \nabla_{\mu}N\nabla^{\nu}\phi^2
-\frac{3\xi\phi^2}{2N}\nabla^2N-\frac{\epsilon}{2N}\left[ 8{\xi}\stackrel{\ast}
{\left(\frac{\phi\stackrel{\ast}\phi}{N}\right)}+3{\omega}
\frac{\stackrel{\ast}\phi\stackrel{\ast}\phi}{N} \right]\nn\\
&&~~~~~+\frac{\epsilon\xi\phi^2}{8N^2}\Bigg[
-3\stackrel{\ast}  g^{\alpha\beta}
 \stackrel{\ast} g_{\alpha\beta}-6g^{\alpha\beta}\stackrel{\ast\ast} g_{\alpha\beta}
-4 \frac{\stackrel{\ast}\phi}{\phi}
g^{\alpha\beta}\stackrel{\ast}g_{\alpha\beta}
+6 \frac{\stackrel{\ast}{N}}{N}g^{\alpha\beta}\stackrel{\ast}g_{\alpha\beta}
\Bigg].\label{potential}
\end{eqnarray}
Substituting the above equation back into Eq. (\ref{constraint1}), we arrive at
\begin{eqnarray}
&&\hspace*{-2em}\frac{\xi \phi^2}{2} R^{(4)}=\frac{\xi}{2}\nabla^2\phi^2-2\frac{\xi}{N} \nabla_{\mu}N\nabla^{\nu}\phi^2
-\frac{3\xi\phi^2}{2N}\nabla^2N+\frac{\omega}{2}g^{\M\N}\na_{\M}{\phi}\na_{\N}\phi-\frac{
 2\epsilon\omega}{N^2}
 \stackrel{\ast}\phi\stackrel{\ast}\phi-\frac{4\epsilon\xi}{N}\stackrel{\ast}
{\left(\frac{\phi\stackrel{\ast}\phi}{N}\right)}\nn\\
&&\hspace*{2em}+\frac{\epsilon\xi \phi^2}{8N^2}  \left(-2 \stackrel{\ast}  g^{\alpha\beta}  \stackrel{\ast} g_{\alpha\beta}+(g^{\alpha\beta}\stackrel{\ast }
g_{\alpha\beta})^2-6g^{\alpha\beta}\stackrel{\ast\ast} g_{\alpha\beta}
+\frac{4\stackrel{\ast}\phi}{\phi} g^{\alpha\beta} \stackrel{\ast}g_{\alpha\beta} +\frac{6\stackrel{\ast}N}{N} g^{\alpha\beta}
 \stackrel{\ast}g_{\alpha\beta}\right).\label{curvature}
\end{eqnarray}

Let us focus on the conformally invariant case with $\omega
=-16\xi/3$ and $V(\phi)=V_0 |\phi|^{\frac{10}{3}}$. In order to find
solutions for this case, we consider the following ansatz:
\begin{eqnarray}
g_{\mu\nu}(x, y)=e^{-2\mu^2(y-y_0)^2} \hat g_{\mu\nu}(x), ~N(x,
y)= N_0e^{-z_1{\mu^2}y^2}, ~ \phi(x,y)=\phi_0
e^{\frac{3}{2}z_2\mu^2y^2}, \label{mga}
\end{eqnarray}
where $N_0$ and $\phi_0$ are constants. For this ansatz, the Eqs.
(\ref{dpotential})$\sim$(\ref{curvature}) become
\begin{eqnarray}
\frac{10}{3}V_0\phi_0^{\frac{7}{3}}&=&e^{2(z_1-z_2)\mu^2
y^2}\frac{\phi_0}{N_0^2}\epsilon\xi\Bigg[80
\left(1-z_2\right)\left(z_1- 1\right)\mu^4 y^2+80zy_0
y(2-z_1-z_2)\mu^4\nonumber\\
&&\hspace*{16em}+40(1-z_2)\mu^2-80y_0^2\mu^4  \Bigg]
\label{master_cc1},\\
V_0\phi_0^{\frac{10}{3}}&=&e^{2(z_1-z_2)\mu^2
y^2}\frac{\phi_0^2}{N_0^2}\epsilon\xi
\Bigg[24\left(1-z_2\right)\left(z_1- 1\right) \mu^4 y^2+24y_0
y(2-z_1-z_2)\mu^4\nonumber\\
&&\hspace*{16em}+12(1-z_2)\mu^2-24y_0^2\mu^4 \Bigg]
\label{master_cc2},\\
R^{(4)}&=&e^{2z_1 \mu^2 y^2}
\frac{2\epsilon}{N_0^2}
\Bigg[24\left(1-z_2\right)\left(z_1- z_2\right) \mu^4 y^2-4y_0
y(z_1-z_2)\mu^4\nonumber\\
&&\hspace*{16em}+12\left (1-z_2  \right)\mu^2
\Bigg]\label{master_cc3}.
\end{eqnarray}
The above equation (\ref{master_cc3}) determines  $R^{(4)}$ as a
function of $y,$ namely each hypersurface $y=\bar{y}$ has different
values of $R^{(4)}$. But, we shall restrict our attention to
four-dimensional Minkowski space. Then, we notice that the
Eqs.(\ref{master_cc1})$\sim$(\ref{master_cc3}) always allow trivial
vacuum solution
 $\phi_0=0,~R^{(4)}=$ arbitrary, independent of $z_1$ and $z_2, $ in general.
The search for nontrivial vacuum with $\phi_0\neq0$ is facilitated
by the fact that the coefficients in Eqs. (\ref{master_cc1}) and
(\ref{master_cc2}) come out right so that the two equations are
identical. Finally, for the 4D Minkowski vacuum ($R^{(4)}=0$), we
can obtain two solutions: $(i)$ the $r.h.s$ of Eqs.
(\ref{master_cc1})-(\ref{master_cc3}) vanishes when $y_0=0$ and
$z_2=1$, which yields $V_0=0,~R^{(4)}=0$ and in this case, $z_1$ is
arbitrary, $(ii)$ for $V_0\neq0$ and $y_0\neq0$, they allow the
solution of $z_1=z_2=1$. We summarize the 4D Minkowski solutions as
follows:
\begin{eqnarray}
&&\nonumber\\
&&\hspace*{-5em}(i)~~~~~g_{\mu\nu}=e^{-2\mu^2y^2}  \eta_{\mu\nu}, ~~N= N_0e^{-z_1{\mu^2}y^2}, ~ \phi=\phi_0
e^{\frac{3}{2}\mu^2y^2},~~V_0=0\\ \label{mga1}
&&\nonumber\\
&&\hspace*{-5em}(ii)~~~g_{\mu\nu}=e^{-2\mu^2(y-y_0)^2}  \eta_{\mu\nu}, ~N= N_0e^{-{\mu^2}y^2}, ~ \phi=\phi_0
e^{\frac{3}{2}\mu^2y^2},~~V_0=-24\epsilon\xi \frac{y_0^2\mu^4}{N_0^{2}\phi_0^{\frac{4}{3}}} \label{mga2}
\end{eqnarray}
\begin{figure}[t!]
\begin{center}
\begin{tabular}{cc}
\includegraphics[width=.8
\linewidth,origin=tl]{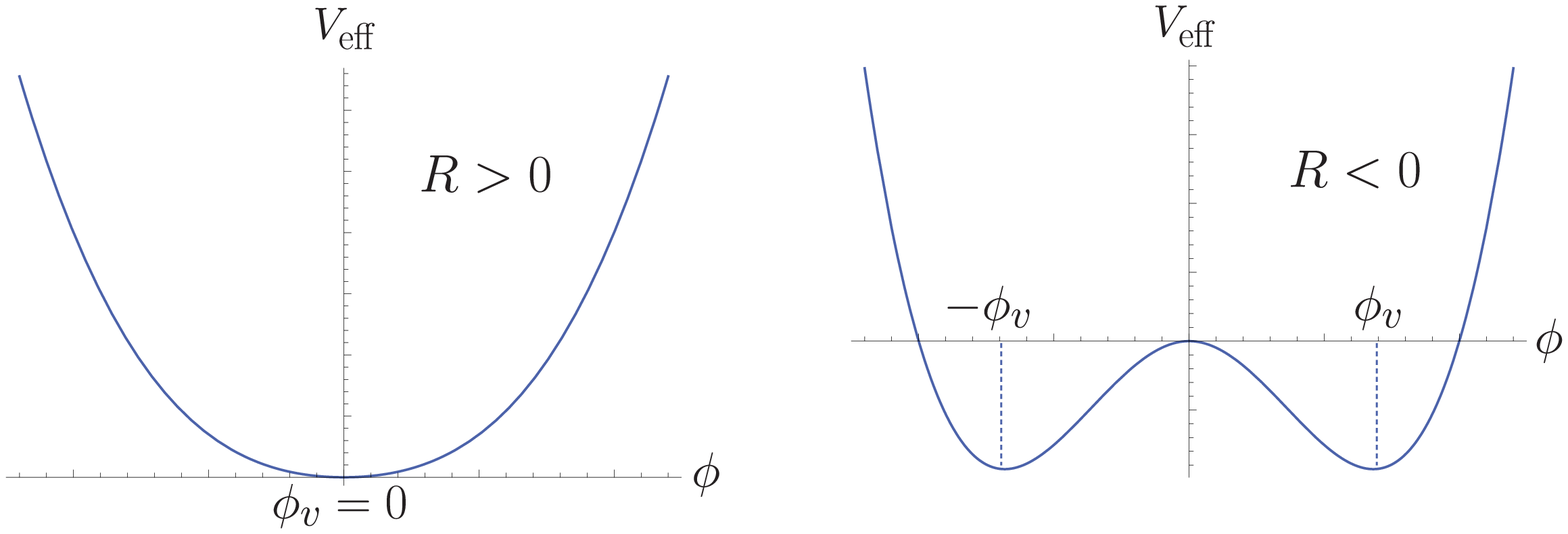}
\end{tabular}
\end{center}
\caption{The effective potential $V_{\rm eff}=-\xi
R\phi^2/2+V_0|\phi|^{\frac{10}{3}}$ with $\xi<0$ and $V_0>0$. The
left panel [unbroken phase] corresponds to the effective potential
with a positive value of the curvature scalar $R$, while the right
one [broken phase] is a case of $R<0$.}
\end{figure}

Now we turn to an issue related to the spontaneous breaking of the
conformal symmetry. We notice that the spontaneous symmetry breaking
can be realized for a negative value of the curvature scalar with
$R<0$ and $V_0>0$. To see this, we consider an effective potential
$V_{\rm eff}$ for the canonical scalar field ($\omega=1$) and
$V_0>0$ in the action (\ref{conac}) as
\begin{eqnarray}
V_{\rm eff}=-\frac{1}{2}\xi\phi^2 R+V_0|\phi|^{\frac{10}{3}}.
\end{eqnarray}
As was mentioned in Eq.(\ref{xiv}), here $\xi$ is fixed as a
negative value of $\xi=-3/16$ for the canonical scalar $\omega=1$,
which preserves the conformal symmetry of the action (\ref{conac}).
Since the solution $(i)$ ($V_0=0$) with a stable equilibrium
$\phi_v=0$ does not provide a symmetry-broken phase, we focus on the
case $(ii)$ with $\epsilon=1$ giving $V_0>0$ (hereafter we fix
$\epsilon=1$).

It turns out that for $(ii)$ with the positive 5D curvature scalar
$R>0$, we have only one vacuum solution of $\phi_{v}=0$, while for
$R<0$, there exist two vacua of $\pm\phi_v$ with non-zero value:
\begin{eqnarray}\label{phiv}
\phi_{v}=e^{\frac{3}{2}\mu^2y_{*}^2}\phi_0,
\end{eqnarray}
where $y_*$ is given by $y_*=4y_0/3\pm\sqrt{3/\mu^2+16y_0^2}/3$. In
this case, the conformal symmetry is spontaneously broken with the
symmetry breaking scale $\phi_v$  given by (\ref{phiv}).
This result is summarized in  Fig.1 which shows that the effective potential with $R>0$ has only one
minimum $\phi_v=0$ ($left$), while for $R<0$, it has two minima with
$\pm\phi_v$ ($right$) which corresponds to the case of spontaneous
breaking of the conformal symmetry.

\section{Tensor perturbation}

In this section, we explore the stability of the solution
$(i),~(ii)$ by performing a tensor fluctuation around the solutions.
Note that the impact of conformal invariance shows up in the
perturbation theory. One can always go to the unitary gauge and
choose $\varphi=0$ in the scalar perturbation with
$\phi=\phi_0+\varphi$. On the other hand, when $\omega/\xi \neq
-16/3,$ it is no longer possible to gauge away the fluctuation, and
$\varphi$ is a dynamical field coupled to other fluctuations.

To investigate a tensor fluctuation around the solutions $(i),(ii),$
we need to consider the tensor perturbation of the metric as
\begin{eqnarray}
ds^2&=&a^2(\eta_{\mu\nu}+h_{\mu\nu})dx^{\mu}dx^{\nu}+N^2 dy^2\\
&=&a^2\Big[(\eta_{\mu\nu}+h_{\mu\nu})dx^{\mu}dx^{\nu}+ dz^2\Big],\label{confm}
\end{eqnarray}
where a new variable $z$ given in a conformally flat metric
(\ref{confm}) satisfies $Ndy=adz$. It is found the equation of
motion for the tensor modes $h_{\mu\nu}$ is given by\begin{eqnarray}\label{mastereq0}
\partial_{z}^2h_{\mu\nu}+A(z)\partial_{z}h_{\mu\nu}-\Box^{(4)}h_{\mu\nu}=0,
\end{eqnarray}
where $h_{\mu\nu}$ satisfy the transverse-traceless gauge conditions ($\partial^{\mu}h_{\mu\nu}=0$,~$h=\eta^{\mu\nu}h_{\mu\nu}=0$) and $A(z)$ is given by
\begin{eqnarray}
A(z)=\frac{3}{a}\frac{\partial a}{\partial
z}+\frac{2}{\phi}\frac{\partial \phi}{\partial z}.
\end{eqnarray}
 Consideration of a separation of variables $h_{\mu\nu}(x,z)=f(z)H_{\mu\nu}(x)$ splits  the equation (\ref{mastereq0})
 into two parts:
\begin{eqnarray}
&&[-\partial_{z}^2+V_{\rm QM}]f= m^2f,\label{mastereq}\\
&&\Box^{(4)}H_{\mu\nu}= m^2H_{\mu\nu}.
\end{eqnarray}
Here $m$ is the mass of four dimensional Kaluza-Klein modes and the corresponding quantum mechanical
potential $V_{\rm QM}$ reads
\begin{eqnarray}
V_{\rm QM}=\frac{1}{2}\frac{\partial A}{\partial z}+\frac{A^2}{4}.
\end{eqnarray}
One can check easily that $V_{\rm QM}$ vanishes for solution $(i)$,
which yields just plane wave solution with constant zero mode. On
the other hand, for solution $(ii)$, it gives an inverse square
potential\footnote{The inverse square potential (\ref{pote})
corresponds to a repulsion from the origin. For an attractive
potential ($\alpha<-1/4$) case with $V(x)=\alpha x^{-2}$, it was
shown in Ref.\cite{Case:1950an} that the quantum mechanical system
has an infinite continuous bound states from negative infinity to
zero.} as
\begin{eqnarray}\label{pote}
V_{\rm QM}=\frac{15}{4z^2},
\end{eqnarray}
where the corresponding Hamiltonian can be written as
\begin{eqnarray}\label{Ham}
H=\frac{1}{2}(p_{z}^2+gz^{-2}) ~~~{\rm with}~ g=\frac{15}{4}.
\end{eqnarray}
It is known in \cite{Calogero:1969xj,Moroz:2009kv} that for the
Schr\"{o}dinger equation (\ref{mastereq}) with the inverse square
potential ($V_{\rm QM}=g/z^2$), the stability of the mode $f$ is
determined by the condition\footnote{This condition yields exactly
the BF bound \cite{Breitenlohner:1982bm} given in the
$d$-dimensional AdS spacetime for the one-dimensional Sch\"{o}dinger
equation:
\begin{eqnarray}
-\partial_x^2\psi+\frac{m^2+\frac{d^2-1}{4}}{x^2}=E\psi
~~\rightarrow~~m^2\ge m_{\rm BF}^2=-\frac{d^2}{4},\nonumber
\end{eqnarray}
when replacing $g$ with $m^2+(d^2-1)/4$.}
\begin{eqnarray}
g>-\frac{1}{4},
\end{eqnarray}
which guarantees that the graviton mode along
the fifth dimension with $g=15/4$ is stable.

Before closing the section, we remark that the graviton mode along
the fifth dimension preserves the residual conformal $SO(2,1)$ symmetry.  It
is well-known that the quantum mechanical system of the Hamiltonian
(\ref{Ham}) is conformally invariant \cite{deAlfaro:1976vlx}, being
reffered to as {\it conformal quantum mechanics} (CQM). To see this,
we first construct the CQM action for $H$ (\ref{Ham}) from the
Lagrangian formalism:
\begin{eqnarray}
S_{\rm CQM}=\frac{1}{2}\int dt \Big(\dot{z}^2-\frac{g}{z^2}\Big),
\end{eqnarray}
which is invariant under the non-relativistic conformal
transformations:
\begin{eqnarray}
t'=\frac{\alpha t+\beta}{\gamma t+\delta},~~~z'=\frac{z}{\gamma
t+\delta}~~~~~{\rm with}~~ ~\alpha\delta-\beta\gamma=1.
\end{eqnarray}
In this case, it is known that three generators, i.e.,  $H$ (time
translation), $D$ (dilatation), $K$ (special conformal) generators
can act with the transformation rules:
\begin{eqnarray}
H;~~~t'=t+\tilde{t},~~~~~~~D;~~~t'={\tilde{d}}^2t,~~~~~~~K;
~~~t'=\frac{t}{\tilde{k}t+1},
\end{eqnarray}
where $\tilde{t},~\tilde{d},~\tilde{k}$ are some constants and the
generators $D$ and $K$ at $t=0$ in addition to $H$ (\ref{Ham}) are
given by
\begin{eqnarray}\label{dko}
D=-\frac{1}{4}(zp_z+p_z z),~~~K=\frac{1}{2}z^2.
\end{eqnarray}
These generators obey $SO(2,1)$ commutations rules given by
\begin{eqnarray}
[H,D]=iH,~~~~~[K,D]=iK,~~~~~[H,K]=2iD,
\end{eqnarray}
whose Casimir invariant ${\cal C}$ is given by ${\cal
C}=(HK+KH)/2-D^2=3/4$.

\section{Summary and discussion}

In this paper, we considered the conformally invariant gravity in
5D, which consists of a scalar field nonminimally coupled  to the
curvature with its potential. We found two solutions $(i)$ and $(ii)$
giving 4D Minkowski vacuuum. By analyzing the dynamics of the metric
perturbations around the solutions, we showed that two solutions are
stable, since the former yields a plane wave solution with the
constant zero mode, whereas the latter gives an inverse square
potential. In particular, it was shown for the solution $(ii)$ that one has unbroken phase when $R>0$, $V_0>0$, while for $R<0$, $V_0>0$ the spontaneous breaking of the conformal symmetry can be realized with the scale  $\phi_v$ given by (\ref{phiv}).

   We point out
that the solution $(ii)$ may lead to a different mechanism
which allows the possibility of a spontaneous breaking of
translational invariance along the extra dimension. To this end,
we consider a solution explicitly to Eq.(\ref{mastereq}) for $m^2=0$
as
\begin{eqnarray}\label{sol12}
f(z)_{m^2=0}=c_1z^{\frac{5}{2}}+c_2z^{-\frac{3}{2}}
\end{eqnarray}
with arbitrary constants $c_{1}$ and $c_{2}$.
The first term of the $r.h.s$ in Eq.(\ref{sol12}) is not normalizable
since the function $f(z)$ diverges at infinity, while the second term can
not lead to a normalizable solution due to its
 divergence at the origin. Thus, there is no normalizable zero mode
 solution. To resolve this problem,
we define a new evolution operator ${\cal R}$
\cite{deAlfaro:1976vlx} given in terms of $K$ (\ref{dko}) and $H$
(\ref{Ham})
\begin{eqnarray}\label{rex}
{\cal R}\equiv\frac{1}{2}\Big(\frac{1}{a}K+aH\Big),
\end{eqnarray}
which yields the eigenvalues of ${\cal R}$ as follows
\begin{eqnarray}
r_n=r_0+n,~~~~~~~r_0=\frac{3}{2}.
\end{eqnarray}
Here, $a$ is some constant with the length dimension. It turns out
that the new evolution operator ${\cal R}$ (\ref{rex}) provides a
normalizable ground state $f_0$:
\begin{eqnarray}\label{fo}
f_0(z)=c_0 z^{\frac{5}{2}} e^{-\frac{z^2}{2}},~~~~z\ge0,
\end{eqnarray}
where a constant $c_0$ is given by $c_0=1$, being obtained from the
normalization condition $\int_{0}^{\infty}|f_0(z)|^2dz=1$.
Importantly, even if we have the normalizable ground state
(\ref{fo}) by introducing the new evolution operator ${\cal R}$
(\ref{rex}), it implements the spontaneous breaking of the conformal
symmetry in the sense that the fundamental length scale $a$ is not
included in the Lagrangian but generated by the particular form of
the vacuum. On the other hand, it should be pointed out that since
the well-defined ground state described by the Hamiltonian $H$ which
generates the
 time-translation  is not present, it may lead a spontaneous symmetry
breaking of time-translational invariance along the dynamical fifth
direction\cite{Fubini:1976jm,D'Hoker:1983ef,Bernard:1983ip,Rabinovici:2007hz}.

We conclude with the following remark.  We see that both  solutions
$(i)$ and $(ii)$ can be characterized by Gaussian warp factor
\cite{Flanagan:2001dy,Alexandre:2008ja,Quiros:2012bh}, where the
maximum value is located at $y=0$ and $y=y_0$, respectively. But,
the vacuum mode $\phi_v$ of Eq. (\ref{phiv}) of the scalar field and
the massless mode of gravity for the broken phase ($R<0$, $V_0>0$)
are not localized on the Gaussian brane, because a value of $y_*$ in
(\ref{phiv}) can not be equivalent to $y_0$ and the zero mode
(\ref{sol12})  is written as $f(y)_{m^2=0}=\tilde{c}_1 e^{-5\mu^2
y_0 y}+\tilde{c}_2 e^{3\mu^2 y_0 y}$ in $y$-coordinate. Thus, it
seems to be hard to describe the brane-world scenario with the
current approach. One possible alternative  to apply our result to
the scenario would be to treat the conformal gravity discussed in
this study (or its variation) as the conformal matter sector and
introduce 5D Einstein-Hilbert action separately. Then, there could
be a possibility of addressing some of the related issues,
especially  the brane stabilization as a consequence of the
spontaneous symmetry breaking.  The  details will be reported
elsewhere.

\section*{Acknowledgments}
This work was supported by  Basic Science
Research Program through the National Research
Foundation of Korea (NRF) funded by the Ministry of
Education (Grant No. 2015R1D1A1A01056572).


\begin{thebibliography}{99}

\bibitem{Weyl:1918ib}
  H.~Weyl,
  Sitzungsber.\ Preuss.\ Akad.\ Wiss.\ Berlin (Math.\ Phys.\ ) {\bf 1918} (1918) 465;

  H.~Weyl,
  Annalen Phys.\  {\bf 59} (1919) 101
   [Surveys High Energ.\ Phys.\  {\bf 5} (1986) 237]
   [Annalen Phys.\  {\bf 364} (1919) 101].
  doi:10.1002/andp.19193641002


\bibitem{Dirac:1938mt}
  P.~A.~M.~Dirac,
  Proc.\ Roy.\ Soc.\ Lond.\ A {\bf 165} (1938) 199.
  doi:10.1098/rspa.1938.0053;

  P.~A.~M.~Dirac,
  Proc.\ Roy.\ Soc.\ Lond.\ A {\bf 333} (1973) 403.
  doi:10.1098/rspa.1973.0070.






\bibitem{Jackiw:2011vz}
  R.~Jackiw and S.-Y.~Pi,
  J.\ Phys.\ A {\bf 44} (2011) 223001.
  doi:10.1088/1751-8113/44/22/223001
  [arXiv:1101.4886 [math-ph]];
 See also  T.~Y.~Moon, J.~Lee and P.~Oh,
  Mod.\ Phys.\ Lett.\ A {\bf 25}, 3129 (2010).
  doi:10.1142/S0217732310034201
  [arXiv:0912.0432 [gr-qc]] and references therein.




\bibitem{Padmanabhan:1986yc}
  T.~Padmanabhan,
  Class.\ Quant.\ Grav.\  {\bf 2} (1985) L105.
  doi:10.1088/0264-9381/2/5/002.


\bibitem{Demir:2004kc}
  D.~A.~Demir,
  Phys.\ Lett.\ B {\bf 584} (2004) 133.
  doi:10.1016/j.physletb.2004.01.044
  [hep-ph/0401163].



\bibitem{Bertolami:2006bf}
  O.~Bertolami and C.~Carvalho,
  Phys.\ Rev.\ D {\bf 74}, 084020 (2006).
  doi:10.1103/PhysRevD.74.084020
  [gr-qc/0607043].


\bibitem{Bertolami:2007dt}
  O.~Bertolami and C.~Carvalho,
  Phys.\ Rev.\ D {\bf 76}, 104048 (2007).
  doi:10.1103/PhysRevD.76.104048
  [arXiv:0705.1923 [hep-th]].

\bibitem{ArkaniHamed:1998rs}
  N.~Arkani-Hamed, S.~Dimopoulos and G.~R.~Dvali,
  Phys.\ Lett.\ B {\bf 429}, 263 (1998).
  doi:10.1016/S0370-2693(98)00466-3
  [hep-ph/9803315];
  L.~Randall and R.~Sundrum,
  Phys.\ Rev.\ Lett.\  {\bf 83}, 3370 (1999).
  doi:10.1103/PhysRevLett.83.3370
  [hep-ph/9905221];
  L.~Randall and R.~Sundrum,
  Phys.\ Rev.\ Lett.\  {\bf 83}, 4690 (1999).
  doi:10.1103/PhysRevLett.83.4690
  [hep-th/9906064];
  P.~D.~Mannheim,
 ``Brane-localized gravity,''
  Hackensack, USA: World Scientific (2005).


\bibitem{Wesson:1999nq}
  P.~S.~Wesson,
  ``Space - time - matter: Modern Kaluza-Klein theory,''
  Hackensack, USA: World Scientific (2007).


\bibitem{PonceDeLeon:2001un}
  J.~Ponce De Leon,
  Mod.\ Phys.\ Lett.\ A {\bf 16} (2001) 2291. 
  doi:10.1142/S0217732301005709
  [gr-qc/0111011].


\bibitem{Doroud:2009zza}
  N.~Doroud, S.~M.~M.~Rasouli and S.~Jalalzadeh,
  Gen.\ Rel.\ Grav.\  {\bf 41} (2009) 2637.
  doi:10.1007/s10714-009-0793-y.

\bibitem{Rasouli:2014dxa}
  S.~M.~M.~Rasouli, M.~Farhoudi and P.~Vargas Moniz,
  Class.\ Quant.\ Grav.\  {\bf 31} (2014) 115002.
  doi:10.1088/0264-9381/31/11/115002
  [arXiv:1405.0229 [gr-qc]].


\bibitem{Rasouli:2016ngl}
  S.~M.~M.~Rasouli and P.~V.~Moniz,
  Class.\ Quant.\ Grav.\  {\bf 33} (2016)  035006.
  doi:10.1088/0264-9381/33/3/035006
  [arXiv:1601.07828 [gr-qc]].






\bibitem{Bars:2013yba}
  I.~Bars, P.~Steinhardt and N.~Turok,
  Phys.\ Rev.\ D {\bf 89}, 043515 (2014).
  doi:10.1103/PhysRevD.89.043515
  [arXiv:1307.1848 [hep-th]].


\bibitem{Cheng:1988zx}
  H.~Cheng,
  Phys.\ Rev.\ Lett.\  {\bf 61}, 2182 (1988).
  doi:10.1103/PhysRevLett.61.2182.


\bibitem{Case:1950an}
  K.~M.~Case,
  Phys.\ Rev.\  {\bf 80}, 797 (1950).
  doi:10.1103/PhysRev.80.797.


\bibitem{Calogero:1969xj}
  F.~Calogero,
  J.\ Math.\ Phys.\  {\bf 10} (1969) 2191.
  doi:10.1063/1.1664820.




\bibitem{Moroz:2009kv}
  S.~Moroz,
  Phys.\ Rev.\ D {\bf 81}, 066002 (2010).
  doi:10.1103/PhysRevD.81.066002
  [arXiv:0911.4060 [hep-th]].


\bibitem{Breitenlohner:1982bm}
  P.~Breitenlohner and D.~Z.~Freedman,
  Phys.\ Lett.\ B {\bf 115} (1982) 197.
  doi:10.1016/0370-2693(82)90643-8.


\bibitem{deAlfaro:1976vlx}
  V.~de Alfaro, S.~Fubini and G.~Furlan,
  Nuovo Cim.\ A {\bf 34} (1976) 569.
  doi:10.1007/BF02785666.


\bibitem{Fubini:1976jm}
  S.~Fubini,
  Nuovo Cim.\ A {\bf 34} (1976) 521.
  doi:10.1007/BF02785664.


\bibitem{D'Hoker:1983ef}
  E.~D'Hoker and R.~Jackiw,
  Phys.\ Rev.\ Lett.\  {\bf 50}, 1719 (1983).
  doi:10.1103/PhysRevLett.50.1719.


\bibitem{Bernard:1983ip}
  C.~W.~Bernard, B.~E.~Lautrup and E.~Rabinovici,
  Phys.\ Lett.\ B {\bf 134} (1984) 335.
  doi:10.1016/0370-2693(84)90011-X.


\bibitem{Rabinovici:2007hz}
  E.~Rabinovici,
  Lect.\ Notes Phys.\  {\bf 737} (2008) 573
  [arXiv:0708.1952 [hep-th]].

\bibitem{Flanagan:2001dy}
  E.~E.~Flanagan, S.~H.~H.~Tye and I.~Wasserman,
  Phys.\ Lett.\ B {\bf 522} (2001) 155.
  doi:10.1016/S0370-2693(01)01261-8
  [hep-th/0110070].



\bibitem{Alexandre:2008ja}
  J.~Alexandre and D.~Yawitch,
  Phys.\ Lett.\ B {\bf 676} (2009) 184.
  doi:10.1016/j.physletb.2009.05.003
  [arXiv:0812.1307 [gr-qc]].



\bibitem{Quiros:2012bh}
  I.~Quiros and T.~Matos,
  arXiv:1210.7553 [gr-qc].


\end{thebibliography}
\end{document}